\documentclass[cite,epsf,12pt]{article}
\usepackage{graphicx}

\newcommand{\beq}{\begin{equation}}
\newcommand{\eeq}{\end{equation}}
\newcommand{\beqarray}{\begin{eqnarray}}
\newcommand{\eeqarray}{\end{eqnarray}}

\begin{document}


\hyphenation{Ginz-burg Le-van-yuk  com-pound multilayered 
ap-pli-ca-tion con-sid-er com-pound su-per-con-duc-ting}

\newcommand{\ie}{{i.e.}}
\newcommand{\eg}{{e.g.}}
\newcommand{\etal}{{\it et al.}}
\newcommand{\eq}[1]{~(\ref{#1})}
\newcommand{\eqdos}[2]{~(\ref{#1},\ref{#2})}
\renewcommand{\epsilon}{\varepsilon}
\newcommand{\lsim}{\stackrel{<}{_\sim}}
\newcommand{\gsim}{\stackrel{>}{_\sim}}

\newcommand{\Tc}{\mbox{$T_c$}}
\newcommand{\Tco}{\mbox{$T_{c0}$}}
\newcommand{\TcH}{\mbox{$T_{c}(H)$}}
\newcommand{\Ds}{\mbox{$\Delta\sigma$}}
\newcommand{\Dchi}{\mbox{$\Delta\chi$}}
\newcommand{\DM}{\mbox{$\Delta M$}}
\newcommand{\Hcdoso}{\mbox{$H_{c2}(0)$}}
\newcommand{\hc}{\mbox{$h^C$}}
\newcommand{\hC}{\hc}
\renewcommand{\epsilon}{\varepsilon}


\newcommand{\titulo}{
High-temperature superconducting fault current microlimiters}

\newcommand{\autor}{J A Lorenzo Fern\'andez$^{1,2}$, M R Osorio$^1$, J A Veira$^1$ and F Vidal$^1$\footnote{Corresponding author ({\tt felix.vidal@usc.es}; fax +34 981531682; tel.~+34 981563100 ext.14031).}}

\newcommand{\direccion}{
$^1$Laboratorio de Baixas Temperaturas e
Superconductividade\\ Departamento de F\'{\i}sica da Materia
Condensada,\\  Universidade de Santiago de Compostela,
E-15782, Spain
\\
$^2$Departamento de F\'isica de la Materia Condensada\\ Facultad de Ciencia y Tecnolog\'ia,\\ Universidad del Pa\'is Vasco,
Bilbao E48080, Spain}

\hrule  
\begin{center}
\mbox{}\vspace{-0.5cm}\\
{\nonfrenchspacing\it 
As~published~in~Supercond.~Sci.~Technol.~{\bf 22},~025009~(2009)\\
}
 \end{center}
  \hrule \mbox{}\\
\mbox{}\vspace{-1cm}\\ 
\begin{center}
  \Large\bf
\titulo\\  \end{center}\mbox{}\vspace{-1cm}\\

\begin{center}\normalsize\autor\end{center} 

\begin{center}\normalsize\it\direccion\end{center}


\mbox{}\vskip0.5cm{\bf Abstract. }
High-temperature superconducting microbridges implemented with YBa$_2$Cu$_3$O$_{7-\delta}$ thin-films are shown to be possible fault current limiters for microelectronic devices with some elements working at temperatures below the superconducting critical temperature and, simultaneously, under very low power conditions (below $1$ W). This is the case in the important applications of superconductors as SQUID based electronics, and technologies for communication or infrared detectors.
In this paper it is shown that the good thermal behavior of these microlimiters allows working in a regime where even relatively small faults induce their transition to highly dissipative states, dramatically increasing their limitation efficiency. The conditions for optimal refrigeration and operation of these microlimiters are also proposed.

\newpage
\setlength{\baselineskip}{18pt}


\section{Introduction}

Fault current limiters (FCL) are one of the most promising applications of high critical temperature ($T_c$) superconductors \cite{Weinstock, Bock2005, Noe2007}. In their resistive version, these devices are directly based on the transition to a dissipative state induced by current densities above $J_c$, the critical current density. Up to now research on FCL based on high-$T_c$ superconductors has focused high power applications, such as electric power distribution networks \cite{Noe2007}. No attention has been paid to their possible use in microelectronic devices with some elements working at temperatures below $T_c$ (the superconducting critical temperature) and, simultaneously, under very low power conditions (below $1$ W). This is the case in the important applications of superconductors as SQUID based electronics and technologies for communication or infrared detectors \cite{Weinstock}. Conventional limiters for electronic applications are usually intended to work at room temperature, or even above, but they are not easily available for cryogenic temperatures, as for instance below $77.3$ K, the boiling temperature of liquid nitrogen under normal pressure. At these low temperatures the performance of semiconductor devices is greatly affected. Moreover, some commercial and widely used low power current limiters, such as those based on polymeric materials, need a long time, of the order of seconds, to react to a fault, this time being longer the lower the fault current \cite{PolySwitch}.
In the above applications the power involved in the faults will remain well below 1 W, which may be easily dissipated in cuprate thin-film bridges having widths in the micrometric range \cite{Vinha2003}.  In addition, it was recently shown that the thermal behaviour of these microbridges may be improved by reducing their relative width \cite{Ruibal2007}. 

The first aim of this paper is to show experimentally that, when used as fault current limiters, it is possible to refrigerate very efficiently the superconducting microbridges and then to work at stationary conditions well above $J_c$, even before the fault. This is a crucial advantage compared with the behavior of the much bigger superconducting thin films needed in high power applications \cite{Weinstock,Noe2007}, because it allows one to appreciably reduce the relative excess voltage needed to trigger the FCL and, simultaneously, dramatically increases their limitation efficiency. The conditions for good refrigeration and for optimal operation are also analyzed, the latter on the basis of the propagating hotspot approach \cite{Skocpol1974, Gurevich1987, Poulin1995}.

\section{Experimental details and thermal requirements}

\begin{figure}
 \begin{center}
  \includegraphics[width=0.55\textwidth]{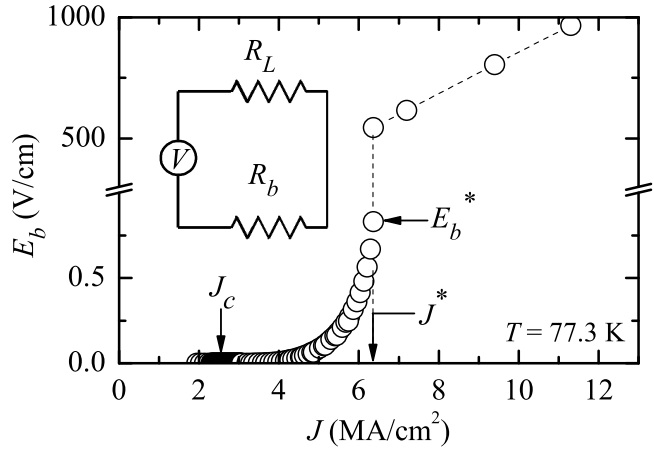}
  \caption{A typical $E-J$ curve obtained at $T=77.3$ K in one of the microbridges used in our experiments (BS7). The inset shows a schematic diagram of the circuit used to probe how the microbridge acts as a FCL, protecting the load resistance, $R_L$, from a voltage fault.}
  \label{fig:figure1}
 \end{center}
\end{figure}

In our experiments, we have used high-quality c-axis oriented YBa$_2$Cu$_3$O$_{7-\delta}$ (YBCO) thin film of thickness 120-300 nm grown on SrTiO$_3$  or sapphire substrates (square shaped 5 mm wide and 0.5-1 mm thick) by using high-pressure on-axis DC sputtering (for those films grown in our laboratory),  or by using pulsed laser deposition (Theva GmbH, Germany). Details about the growth parameters, characterization of the films and preparation of the microbridges have been described elsewhere \cite{Vinha2003}. A typical electric field versus current density ($E-J$) curve, corresponding to one of the microbridges (denoted BS7) used in our experiments, is shown in figure \ref{fig:figure1}. This microbridge, with length, width and thickness 385 $\mu$m, 28 $\mu$m and 300 nm, respectively, and with $T_c = 88.6$ K, has been grown on a sapphire substrate. The two characteristic current densities are indicated in this figure: the critical, $J_c$, at which dissipation first appears, and the so-called supercritical, $J^*$, at which the microbridge is triggered into highly dissipative states. This $V-I$ behavior makes these microbridges good candidates as FCL, but their usefulness also crucially depends on their thermal stability before and during the fault and their thermal reversibility once the fault disappears. 

Some estimations may illustrate how the practical method of fulfilling the thermal requirements indicated above depends very much on the relative dimensions of the superconducting bridges and their substrates. Note first that at temperatures between 80 K and $T_c$ the heat transfer coefficient between the YBCO bridges and their substrates is of the order of $h_{bs} \simeq 10^3$ W/cm$^2$K, whereas towards the N$_2$ liquid or between the substrate and its copper holder ($h_{sr}$) they are typically $1000$ times less \cite{Mosqueira1993,Duron2007}. These values already indicate that the thermal exchanges of the bridge will mainly be through its substrate. In addition, if the corresponding Biot number \cite{Chapman}, $Bi$, is much less than one and, simultaneously, the thermal diffusion length, $L_{th}$, is larger than the corresponding thickness, \emph{e}, the increase in the microbridge temperature relative to that of the bath, $\Delta T_b$, may be crudely approximated as,
\begin{equation}
	\label{equation1}
   \Delta T_b = e_b J E \left( \frac{1}{h_{bs}} + \frac{A_b}{A_s} \frac{1}{h_{sr}} \right){,}
\end{equation}
where $e_b$  is the microbridge thickness and $A_b$ and $A_s$ are the surface areas of the superconductor bridge and, respectively, its substrate. As $h_{bs} \gg h_{sr}$, $\Delta T_b$  will directly depend on  $A_b/A_s$, the condition of ``thermal smallness'' being, 
\begin{equation}
	\label{equation2}
   \frac{A_b}{A_s} < \frac{h_{sr}}{h_{bs}},
\end{equation}
under which $\Delta T_b$ approaches its ``intrinsic'' value, which is the temperature increase just associated with the effective thermal resistance of the bridge-substrate interface.

For the BS7 microbridge and under faults with characteristic times above 10 ms (of the order of commercial ac current periods), the conditions, $Bi \ll 1$ and  $L_{th} > e$ fully apply. In addition, as $A_b/A_s \approx 4 \times 10^{-4}$  whereas $h_{sr}/h_{bs} \approx 10^{-3}$,  this microbridge is ``thermally small''. One may use equation (\ref{equation1}) to estimate $\Delta T_b$ when the microbridge is in stationary conditions under currents just below $J^*$. From figure \ref{fig:figure1}, the power density involved is around $5 \times 10^6$ W/cm$^3$ which leads, by also using the appropriate parameter values\footnote{The substrate thermal conductivities ($K_s$) and heat capacities ($C_s$) are of the order of 10 W/cm K and 0.6 J/cm$^3$ K for sapphire, and 0.2 W/cm K and 1 J/cm$^3$ K for SrTiO$_3$, whereas for YBCO bridges $K_b\approx 0.05$ W/cm K and $C_b \approx 0.7$ J/cm$^3$ K. See, e. g., Refs. \cite{Mosqueira1993, Duron2007, Chapman}.}, to $\Delta T_b \approx 0.1$ K. Even for faults involving powers ten times higher than those considered above, $\Delta T_b$ will remain below 1 K. One may also use equation (\ref{equation1}) to roughly estimate that if the superconducting bridge had $A_b/A_s$  $1000$ times larger, as it is the case of those currently proposed for high power applications \cite{Weinstock,Noe2007},  $\Delta T_b$ would take values at least two orders of magnitude higher. To confirm these values at a quantitative level, $\Delta T_b$ has been calculated by using a finite element method similar to the one described in \cite{Maza2008}. Under the same conditions as above, for the BS7 microbridge we found again $\Delta T_b \approx 0.1$ K.

In the case of SrTiO$_3$ substrates, which have relatively poor thermal conductivities, the condition $Bi\ll1 $ no longer applies. Therefore, a term proportional to $1/k_s$ (i.e. to the inverse of the thermal conductivity of the substrate) must be added in equation \ref{equation1}. For microbridges under the same conditions as before, this leads to  $\Delta T_b \approx 2\;$K, a value that is confirmed by using the finite element method mentioned above. Let us stress, finally, that for a bridge on sapphire but having a surface relative to that of its substrate $1000$ times larger, the finite element method yields $\Delta T_b \approx 50$ K.

\section{Experimental results: optimal operation}

\begin{figure}
 \begin{center}
  \includegraphics[width=0.55\textwidth]{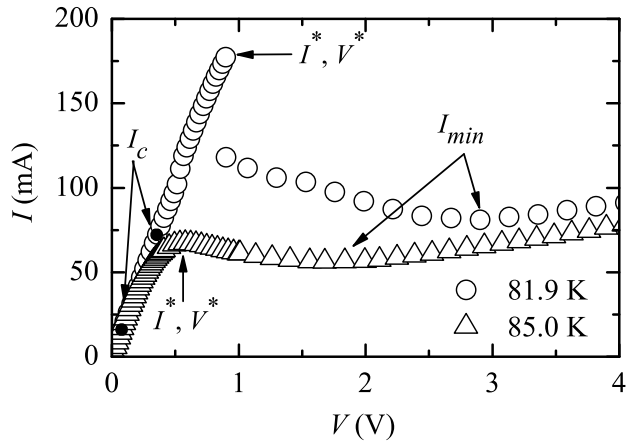}
  \caption{Two $I-V$ curves, showing their strong bath temperature dependence, of the $R_L-R_b$ circuit schematized in the inset of figure \ref{fig:figure1} In these examples $R_b$ is the microbridge BS7 and $R_L=4.9\;\Omega$.}
  \label{fig:figure2}
 \end{center}
\end{figure}

To probe a microlimiter with low thermal dimensions, we have implemented the electrical circuit schematized in the inset of figure \ref{fig:figure1}, with the microbridge BS7 as $R_b$  connected in series to the variable load resistance $R_L$, this last one representing the impedance of the circuit to be protected. The measurements were made in a cryostat with the sample submerged in a forced flow of helium gas. The temperature of the copper holder of the microlimiter was measured with a platinum thermometer and regulated with an electronic system which ensures a temperature stabilization better than 0.05 K. Two examples of the $I-V$ curves obtained in this $R_L-R_b$  circuit (with $R_L =4.9\;\Omega$, this value taking already into account the resistance of the circuit electrical wires, of the order of $1.9\;\Omega$) by using the electronic system described elsewhere \cite{Vinha2003,Ruibal2007} are shown in figure \ref{fig:figure2}. In these curves the voltage was imposed and acquired during pulses of 1 s,  a time much longer than the one needed by the microlimiter to reach the stationary state. The bath temperatures were 81.9 K (circles) and 85.0 K (triangles). As below $V^*$ the flux-flow resistance of the microbridge remains much lower than $R_L$, both curves are almost linear up to $V^*$.  Together with the low heating  estimated above, this quasi-ohmic behavior is crucial to allow the microlimiter to work just below $V^*$  under stationary conditions and without disturbing the circuit to be protected. 

The results presented in figure \ref{fig:figure2} also illustrate two other central aspects of the microbridges when working as FCL.  Note first that, once the source voltage overcomes $V^*$, the current in the circuit varies quite slowly even up to voltage faults as important as four times $V^*$, the current taking a minimum value, $I_{min}$, at some (temperature-dependent) voltage. Moreover, the sharp drop at $V^*$ of the current is also temperature dependent, being almost absent in the curve at 85.0 K. Both aspects are related and may be explained in terms of the approaches based on the propagation of self-heating hotspots \cite{Skocpol1974,Gurevich1987,Poulin1995}:  Above $V^*$ part of the microbridge becomes normal and then, as the total voltage of the circuit is fixed, the increase in resistance gives rise to a decrease in current up to the minimum current, $I_{min}$, capable of sustaining the normal zone. If the fault voltage increases, the length of the hotspot will grow accordingly, keeping the current roughly constant. At a quantitative level, both aspects may be easily explained by just taking into account the reduced temperature ($T/T_c$) dependence of $I^*$ and $I_{min}$,
\begin{equation}
   \label{equation3}
	I(T) = I (0) \left( 1- \frac{T}{T_c} \right)^n\;,
\end{equation}
with $n= 3/2$  for  $I^*$ \cite{Vinha2003,Ruibal2007} and $1/2$ for $I_{min}$ \cite{Skocpol1974,Gurevich1987,Poulin1995}. Therefore, if the reduced temperature increases, both the discontinuity at $V^*$ and the ratio $I^*/I_{min}$ will decrease, in agreement with the results of figure \ref{fig:figure2}. The ``optimal'' reduced temperature for the microlimiter operation, $T_{op}/T_c$, will be then given by the condition $I^*( T_{op}) = I_{min} (T_{op})$. By using equation (\ref{equation3}), this leads to,
\begin{equation}
   \label{equation4}
	\frac{T_{op}}{T_c} = 1 - \frac{I_{min}(0)}{I^*(0)}\; .
\end{equation}
At $T_{op}$  the current limited during the voltage fault will be roughly equal to the nominal one. As $I_{min}(0)< I^*(0)$,  $T_{op}$ will be near, but below enough, $T_c$ to make $I^*(T_{op})$ adequate for the practical operation of the microlimiter under such an optimal temperature.

\begin{figure}
 \begin{center}
  \includegraphics[width=0.55\textwidth]{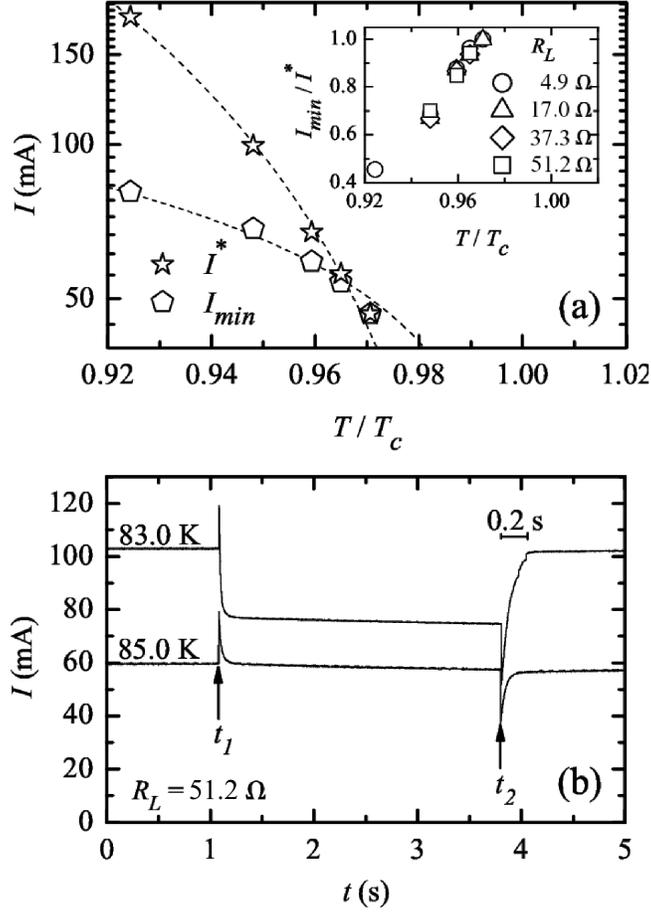}
  \caption{(a) Reduced temperature dependence of $I^*$ and $I_{min}$ in the same $R_L-R_b$ circuit as in figure \ref{fig:figure2}, but with $R_L = 51.2\;\Omega$. The dashed lines are least squares fits of the corresponding theoretical expressions. These curves cross at $T_{op}$. The inset shows the reduced temperature dependence of $I_{min} / I^*$ for various load resistances. (b) Two examples of the time evolution of the current measured in the presence of voltage faults between $t_1$ and $t_2$. See main text for details.}
  \label{fig:figure3}
 \end{center}
\end{figure}

The data of figure \ref{fig:figure3}(a) for the microbrige BS7 confirm the results stressed above.  The dashed lines are least squares fits of equation (\ref{equation3}), with the corresponding critical exponents, to the experimental data points. This leads to $I_{min}(0) \approx 290$  mA and $I^*(0) \approx 8540$ mA and then $T_{op}/T_c \approx 0.97$,  as observed in figure \ref{fig:figure3}(a). Although the corresponding  $T_{op} \approx 85.5$ K is indeed very close to $T_c$,  $I^*(T_{op})$  is still around 50 mA, the right order of magnitude for many potential applications \cite{Weinstock}. Some data on the reduced temperature dependence of $I^*/I_{min}$, always corresponding to BS7 but with different load resistances, are presented in the inset of figure \ref{fig:figure3}(a). These last results show that $I^*(T)/I_{min}(T)$ is indeed independent of the load resistance and that it approaches unity when the working temperature approaches some value just below $T_c$.

The results summarized in figure \ref{fig:figure3}(b) illustrate how a superconducting microlimiter protects a circuit similar to the one of the inset of figure \ref{fig:figure1} from voltage faults. In these examples, $R_S$ is again the BS7 microbridge and $R_L = 51.2\;\Omega$ (taking into account the resistance of the circuit electrical wires). For the curve measured at 83.0 K, the applied voltage was 5.8 V before ($t_1$) and after ($t_2$) a fault regime of 9.5 V. For the curve at 85.0 K, which is near $T_{op}$, these values were 3.5 V and 7.5 V, respectively. As expected, the protection is excellent, whereas there is an overprotection when working well below $T_{op}$. In both cases the recovery after the fault is achieved under current. This is another considerable practical advantage when compared to the superconducting limiters used in high power applications \cite{Weinstock,Noe2007}. These results suggest that optimal current limitation in superconducting electronics could be accomplished by their own conductive pathways after a proper design (e. g. by decreasing the width of the pathway at well-refrigerated selected locations), thus improving compactness.

\section{Conclusions}
High-temperature superconducting microbridges implemented with YBa$_2$Cu$_3$O$_{7-\delta}$ thin films have been shown to be possible fault current limiters for microelectronic devices with some elements working at temperatures below the superconducting critical temperature and, simultaneously, under very low power conditions (below $1$ W). Our results demonstrate experimentally that by using substrates of high thermal conductivity, for instance sapphire, these microbridges may work in a regime where even relatively small faults induce their transition to highly-dissipative states, which dramatically increases their limitation efficiency. The criteria for good refrigeration and for optimal operation have been also obtained. Our results suggest that these superconducting microlimiters may be particularly useful in the important applications of low and high $T_c$ superconductors as SQUID or infrared detectors. In some applications this type of fault current superconducting microlimiters could be achieved by just decreasing the width of the superconducting current leads at well-refrigerated selected locations.

\mbox{}

\mbox{}

\mbox{}\\ {\Large \bf Acknowledgements}\\ \mbox{}\\
\nonfrenchspacing
This work has been supported under Projects No. 2006/XA049 and No. 07TMT007304PR (XUGA-FEDER) and by Project. No. FIS2007-63709 (MEC-FEDER). J. A. L. F. acknowledges financial support from the Basque Government, through a grant from Education, Universities and Research Department.

\newpage

\end{document}